\documentclass[proceedings]{rmaa}

\usepackage{rmaacite}

\def\tento#1{$\times$10$^{#1}$}
\def\degr{\hbox{$^\circ$}}
\def\arcmin{\hbox{$^\prime$}}
\def\arcsec{\hbox{$^{\prime\prime}$}}
\def\HII{{\rm H}$\;\!\!${\sc ii}\,\,}
\def\HI{{\rm H}$\;\!\!${\sc i}\,\,}

\title{ROSAT X-ray observations of Central Galactic Starbursts}

\author{D.~Tsch\"oke and G.~Hensler
        \affil{Institute of Theoretical Physics and Astrophysics,
               University of Kiel, Germany}}

\fulladdresses{
\item D.~Tsch\"oke and G.~Hensler: Institute of Theoretical Physics 
      and Astrophysics, University of Kiel, Leibnizstr. 15,
      D-24098 Kiel, Germany (tschoeke,hensler@astrophysik.uni-kiel.de)}

\shortauthor{Tsch\"oke \& Hensler}
\shorttitle{X-ray observations of Central Starbursts}

\abstract{
We present spatial and spectral results of ROSAT observations of two galaxies: NGC~4569 and NGC~4303. Both objects are members of the Virgo Cluster and have nearby smaller companions with which they have probably gone through an encounter.

Both soft X-ray spectra (0.1-2.4 keV) can be fitted with a two-component model consisting of a power-law component from a non-thermal emission of a compact active nucleus and a Raymond-Smith model for a hot thin thermal plasma originating from starbursts. The X-ray luminosities are in the range of  $10^{40}$ erg s$^{-1}$. The central source of NGC~4569 is very compact and can not be resolved with the ROSAT HRI detector, while there is extended soft X-ray emission in the galactic disk as well as above it. NGC~4303 is dominated by the active nucleus in the X-ray. But several additional disk sources associated with \HII regions can be observed.
}


\begin{document}

\maketitle

\section{Introduction}
\label{intro}

The phenomenon of galactic nuclear activity is strongly connected with starbursts, especially in galaxies with low luminous active nuclei, like LINERs and Seyfert galaxies. Beside UV and optical observations the analysis of X-ray data reveals composite spectra, indicating non-thermal nuclear activity and soft thermal emission from the hot gas of a central starburst. One main aspect of interest is the triggering mechanism of galactic activity in different morphologies, like interacting systems \cite{Tsc99} and bars.

Here we investigate luminosity and spectral distribution of the soft X-ray emission from two nearby barred galaxies with nuclear starburst activity. In both nearby galaxies The X-ray emission can be clearly distinguished between different galactic components: the center, the disk, and, in the case of NGC~4569, the halo. Both galaxies have been classified as LINERS from optical spectroscopy (, NGC~4569: \pcite{Wil85}; NGC~4303: \pcite{Huc82}).

NGC~4569 is a bright early-type spiral in the Virgo Cluster, one of the few blue-shifted galaxies outside the local group. It is gas-deficient in the outer spiral arms, the neutral hydrogen content is strongly concentrated in the inner region \cite{Cay90}. The bright nucleus, embedded in a normal stellar bulge, is probably the result of a recent star formation episode \cite{Sta86}.

NGC~4303 is a barred late-type spiral galaxy and also member of the Virgo Cluster. Indications for a high star formation rate in NGC~4303 are the high density of \HII regions \cite{Hod83,Mar92} and three observed supernovae \cite{Dyk92}. It shows also strong radio emission distributed over the entire disk \cite{Con83}. Two nearby galaxies, NGC~4303\,A \cite{Con83} and NGC~4292 \cite{Cay90}, are possible candidates for a past interaction with NGC~4303.

\section{Spatial distribution, X-ray spectra and luminosities}

\subsection{NGC~4569}

The global X-ray spectrum of NGC~4569 can be described by a two-component fit consisting of a power-law component (predominantly from the nucleus) and a thermal plasma component (hot gas from SNRs and superbubbles). The resulting value of hydrogen absorption is in agreement with the Galactic foreground, supporting the observed \HI deficiency \cite{Cay90}. The total luminosity in the ROSAT band for NGC~4569 amounts to $\sim$2\tento{40} erg s$^{-1}$.

Two interpretations for the emission of the compact nucleus in the HRI image seem possible: either X-ray emission from an unresolved compact nuclear source, to be associated with the LINER-type nucleus of NGC~4569 in the optical, or a slightly extended nuclear starburst region, similar to the one found in NGC~1808 \cite{Jun95} at a scale of $\sim$1~kpc. The spectrum of the nuclear source can well be fitted with a power law spectrum - with the spectral resolution of the ROSAT PSPC it is not possible to clearly determine the nature of the central source - either the contribution from high-mass binaries in the central starburst, or an AGN component with a spectral distribution in the ROSAT band which could be described by a power-law fit. 

X-ray emission from the disk of NGC~4569 is clearly visible in the PSPC 
image (Fig.~1). The total disk emission in X-rays from
either side (NE vs. SW) is comparable, Their spectral energy distribution
however differ significantly. The contribution from 
the northern part is significantly harder. This may indicate 
either higher absorption in the north, or an intrinsically harder X-ray 
spectrum caused by the cumulative impact of several SNRs or high-mass X-ray
binaries. 
Higher internal absorption towards the northern part of NGC~4569 seems 
difficult to explain since the \HI content of that galaxy is strongly confined
to the central part \cite{Cay90}.

NGC~4569 has an inclination of $\sim$65\degr.  The galaxy 
appears extended in soft X-rays,
the X-ray emission in the ROSAT band shows contributions from areas
outside the nucleus and outside the disk.  In contrast to
edge-on galaxies, where emission components from disk and halo can be 
quite easily distinguished, it is more difficult 
to disentangle these components in the case of NGC~4569.  

The overall distribution of the soft X-ray emission appears to be 
very asymmetric with a spur extending from the 
galaxy center and some additional diffuse patches west of the 
galaxy. The source northwest of the galactic center is probably not connected to NGC~4569. A further source southwest of the nucleus of NGC~4569 can be identified with a background galaxy visible in the optical. We derived a 
hardness ratio (0.5-2.0~keV vs. 0.1-0.4~keV) of -0.02$\pm$0.17 for 
the area west of NGC~4569, excluding both these independent X-ray sources.  The total number of counts after background subtraction for this area
is only 68, hence no meaningful spectrum could be derived.
The hardness ratio is compared with the results from model spectra.
A value of HR = 0.0 results for a thermal plasma spectrum
(e.g. \pcite{Ray77}) with a plasma temperature of 0.2~keV.  
Hence, the X-ray emission from the area outside the disk can be explained
with hot gas of a temperature of approximately 2\tento{6}~K.

An additional argument in favour of diffuse X-ray emission in the halo 
of NGC~4569 comes from the very good spatial correlation of the 
soft X-ray (0.1-0.4 keV) emission with a large diffuse H$\alpha$ spur (Fig.~1). Similar correlations were observed in a number of nearby 
starburst galaxies (e.g. M82: \pcite{Sho98}, NGC~3628: \pcite{Dah96}, 
NGC~4666: \pcite{Dah97}) and are interpreted as a signature for 
an outflow of hot gas from the starburst center.  Our X-ray 
data are fully consistent with such an interpretation in the case of 
NGC~4569.

\subsection{NGC~4303}

Obvious differences between NGC~4303 and NGC~4569 are that (1) NGC~4303 has a much lower inclination (27\degr; \pcite{Guh88}) and (2) the amount of \HI is more distributed over the total galactic disk and much higher in the outer parts than in NGC~4569, which is closer to the Virgo Cluster center. Consequently, the first point allows better observation possibilities to distinguish between disk and central source. Vice versa, it is not possible to observe a soft halo component above the galactic disk, as can be seen in NGC~4569 or in more details in edge-on starburst galaxies (e.g. \pcite{Vog95}).

Another difficulty in the X-ray observations of NGC~4303 is that no spatial distribution can be obtained from the PSPC data due to the fact that the galaxy is 17\arcmin\ off-axis in the field of view. This degenerates the spatial resolution of the observation to $\sim$67\arcsec.

The global X-ray spectrum of NGC~4303 can be described by a similar fit to that of NGC~4569 with a non-thermal power-law component and a Raymond-Smith component. The resulting value of hydrogen absorption is with 3.4\tento{20} cm$^{-2}$ by a factor of 2 higher than the Galactic foreground. This in agreement with the distribution of \HI gas over the whole galactic disk of NGC~4303. The total luminosity of NGC~4303 in the ROSAT band is $\sim$5\tento{40} erg s$^{-1}$.

The HRI image reveals a dominant central source and several additional sources distributed along the spiral arms (Fig.~2). UV observations \cite{Col99} result in an unresolved ($\le$8 pc) active nucleus (compact stellar cluster or AGN) and a surrounding spiral-shaped region of very young massive star forming clusters (2--3 Myr). This picture leads to attribute a power-law component to the nuclear source, contributing more than 80\% to the total spectrum. This is in agreement with the results from the HRI observation, where the main part originates from the center of the galaxy. The coincidence of the disk sources with \HII regions in the spiral arms is in favour of several star forming regions with SNRs to be the origin of the observed X-ray emission. The sources at the ends of the bar are in agreement with expected gas flows and accumulation within the bar. The eastern spiral feature in the optical and H$\alpha$ with the X-ray source at its bend suggest a past tidal disturbence in the gas dynamics leading to enhanced star formation.

Further contribution to the X-ray flux from HMXBs in the disk sources and from a circumnuclear starburst at a distance up to $\sim$1 kpc from the center can not be excluded from the HRI observation. 

\section{Conclusions}

A comparison of the total luminosities of NGC~4569 and NGC~4303 in the soft X-ray with a larger sample of normal and active galaxies \cite{Gre92} shows good agreement with the galaxies (including LINERs and starburst galaxies) in that sample. Most of the soft X-ray emission in both galaxies comes from their nuclear regions, but additional components from the disk and, in the case of NGC~4569, possibly from the halo, can be distinguished. The existence of a central superwind in NGC~4569 is supported by two facts: First, the extended emission component is predominantly at one side coincident with an H$\alpha$ feature at high z. This one-sided geometry would be expected in an ambient medium with a density gradient along the z-direction and a starburst region slightly above the galactic plane. Second, its spectral distribution (hardness ratio) can be explained with a hot plasma of a temperature of a few 10$^6$ K, which would be expected in the halo of starburst galaxies.

The nature of the nuclei in NGC~4569 and NGC~4303 is still unknown. X-ray images and spectra indicate a coexistence of a dominating AGN and a starburst in both galaxies. Both harbour a compact central source, emitting most of the soft X-ray. Nevertheless, massive nuclear stellar clusters with enhanced star formation can not be excluded to account for the sources.

\begin{figure}
\vspace*{-3cm}
\includegraphics[width=13cm,angle=-90]{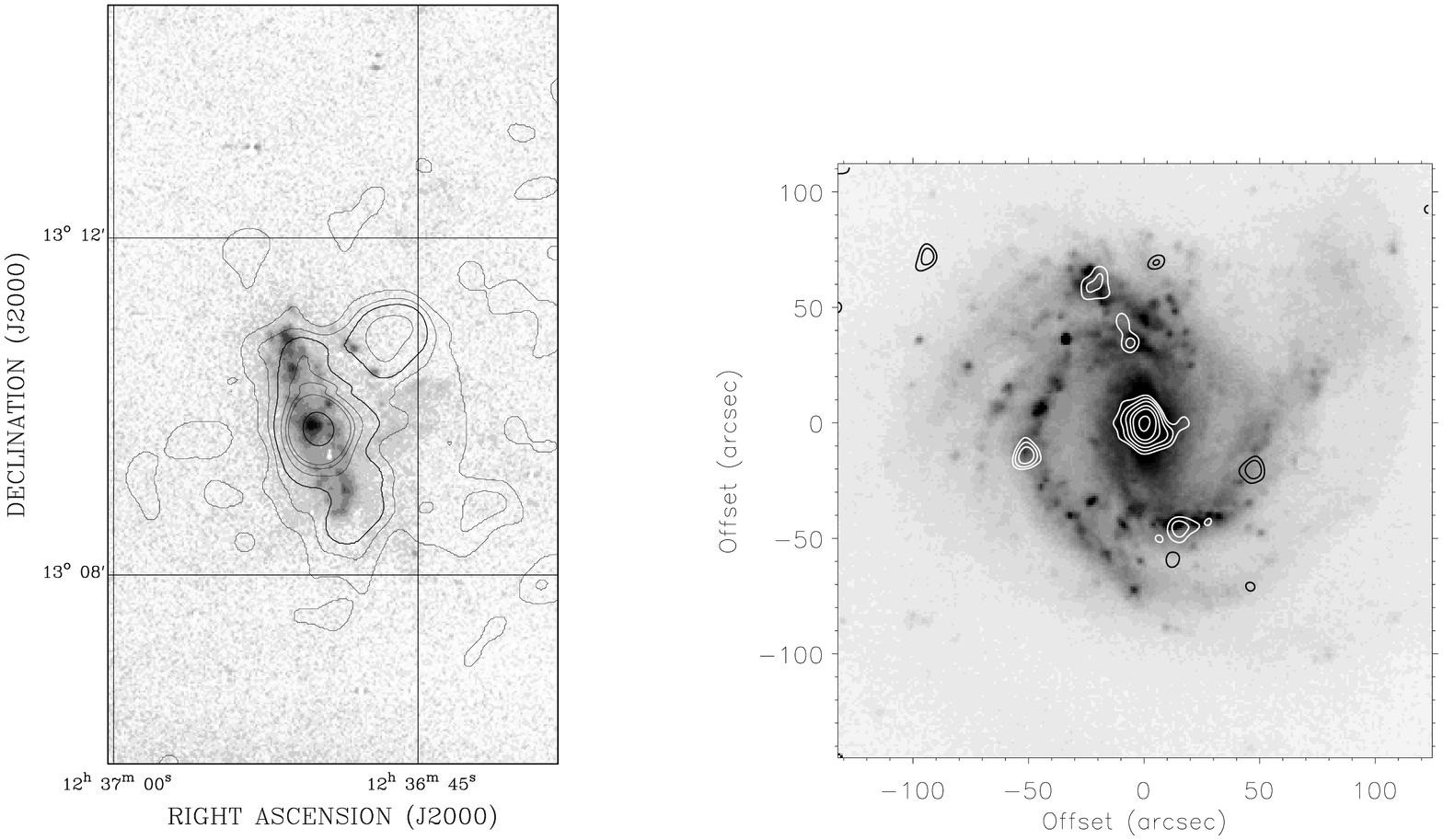}
\parbox{7.5cm}{Figure 1. NGC~4569: PSPC X-ray contours over H$\alpha$ image.}
\hfill
\parbox{7.5cm}{Figure 2. NGC~4303: HRI X-ray contours over R band image \cite{Fre96}.}
\end{figure}

\end{document}